\documentclass[onecolumn,english,aps,prl,nopacs,amsmath,amssymb,superscriptaddress]{revtex4}
\usepackage{graphicx}
\usepackage{amssymb}
\usepackage{natbib}
\usepackage{color}
\usepackage{amsmath}

\newcommand{\BFA}{BaFe$_2$As$_2$}

\begin{document}

\title{In-plane uniaxial pressure-induced out-of-plane antiferromagnetic 
moment and critical fluctuations in {\BFA}}

\author{Panpan Liu}
\thanks{These authors made equal contributions to this work.}
\affiliation{Center for Advanced Quantum Studies and Department of Physics, Beijing Normal University, Beijing 100875, China}
\author{Mason L. Klemm}
\thanks{These authors made equal contributions to this work.}
\affiliation{Department of Physics and Astronomy,
Rice University, Houston, Texas 77005, USA}

\author{Long Tian}
\affiliation{Center for Advanced Quantum Studies and Department of Physics, Beijing Normal University, Beijing 100875, China}
\author{Xingye Lu}
\email{luxy@bnu.edu.cn}
\affiliation{Center for Advanced Quantum Studies and Department of Physics, Beijing Normal University, Beijing 100875, China}

\author{Yu Song}
\affiliation{Department of Physics, University of California, Berkeley, California 94720, USA}
\author{David W. Tam}
\affiliation{Department of Physics and Astronomy,
Rice University, Houston, Texas 77005, USA}

\author{Karin Schmalzl}
\affiliation{Forschungszentrum J$\ddot{u}$lich GmbH, J$\ddot{u}$lich Centre for Neutron Science at ILL, 71 avenue des Martyrs, 38000 Grenoble, France}

\author{J. T. Park}
\affiliation{Heinz Maier-Leibnitz Zentrum (MLZ), Technische Universit$\ddot{a}$t M$\ddot{u}$nchen, 85748 Garching, Germany}

\author{Yu Li}
\affiliation{Department of Physics and Astronomy,
Rice University, Houston, Texas 77005, USA}

\author{Guotai Tan}
\affiliation{Center for Advanced Quantum Studies and Department of Physics, Beijing Normal University, Beijing 100875, China}

\author{Yixi Su}
\affiliation{J$\ddot{u}$lich Centre for Neutron Science, Forschungszentrum J$\ddot{u}$lich GmbH, Outstation at MLZ, D-85747 Garching, Germany}

\author{Fr$\rm \acute{e}$d$\rm \acute{e}$ric Bourdarot}
\affiliation{University Grenoble Alpes, CEA, IRIG, MEM-MDN Grenoble, France
}
\author{Yang Zhao}
\author{Jeffery W. Lynn}
\affiliation{NIST Center for Neutron Research, National Institute of Standards and Technology, Gaithersburg, Maryland 20899, USA}

\author{Robert J. Birgeneau}
\affiliation{Department of Physics, University of California, Berkeley, California 94720, USA}
\affiliation{Materials Sciences Division, Lawrence Berkeley National Laboratory, Berkeley, California 94720, USA}
\affiliation{Department of Materials Science and Engineering, University of California, Berkeley, California 94720, USA}

\author{Pengcheng Dai}
\email{pdai@rice.edu}
\affiliation{Department of Physics and Astronomy,
Rice University, Houston, Texas 77005, USA}

\date{\today}

\begin{abstract}
A small in-plane external uniaxial pressure has been widely used as an effective method to acquire single domain iron pnictide {\BFA}, which exhibits twin-domains without uniaxial
strain below the tetragonal-to-orthorhombic structural (nematic) 
transition temperature $T_s$.
Although it is generally assumed that such a pressure will not affect 
the intrinsic electronic/magnetic properties of the system, it is known to 
 enhance the antiferromagnetic (AF) ordering temperature $T_N$ ($<T_s$) and  
create in-plane resistivity anisotropy above $T_s$.  
Here we use neutron polarization analysis to show that  
such a strain on {\BFA} also induces a static or quasi-static out-of-plane ($c$-axis) AF order and its associated critical spin fluctuations near $T_N/T_s$. Therefore, uniaxial pressure necessary to detwin single crystals of {\BFA} actually rotates the easy axis of the collinear AF order
near $T_N/T_s$, and such effect due to spin-orbit coupling must be taken into account to unveil the intrinsic electronic/magnetic properties of the system.  
\end{abstract}

\maketitle

Understanding the intrinsic electronic, magnetic, and nematic properties of iron pnictides  
such as BaFe$_2$As$_2$ form the basis to unveil the microscopic origin of high-temperature superconductivity because the system is a parent compound of iron-based superconductors \cite{hosono,Johnston,stewart,scalapinormp,dai}. 
As a function of decreasing temperature, {\BFA} first exhibits a tetragonal-to-orthorhombic structural transition at $T_s$ and forms a nematic ordered phase, followed closely by a collinear antiferromagnetic (AF) order with moment along the $a$-axis of the orthorhombic lattice 
below the N$\rm \acute{e}$el temperature $T_N$ ($\leq T_s$) [Fig. 1(a)] 
\cite{qhuang,kim2011,RMFernandes2014,AEBohmerCRP}. Since single crystals of {\BFA} form twin-domains in the orthorhombic state below $T_s$, an external uniaxial pressure 
applied along one-axis of the orthorhombic lattice 
 has been widely used as an effective method to acquire 
single domains of iron pnictide crystals and determine their  
intrinsic transport \cite{fisher,JHChu2010,JHChu2012,Tanatar10,Man15,Tam19}, electronic \cite{MYi2017,Pfau2019,Watson2019}, and magnetic \cite{Lu14,MQHe,xylu18} properties [inset in Fig. 1(b)]. Although uniaxial 
pressure necessary 
to detwin single crystals of {\BFA} is known to increase $T_N$ [Fig. 1(b)] \cite{Dhital2012,Dhital2014,YSong2013,Tam2017} and create in-plane resistivity anisotropy above $T_s$ \cite{Man15}, it is generally assumed that it only induces a small strain on the sample and does not significantly modify the electronic and magnetic properties of the system \cite{fisher,JHChu2010,JHChu2012,Tanatar10,Man15,Tam19,MYi2017,Pfau2019,Watson2019,Lu14,xylu18}. Recently, nuclear magnetic resonance (NMR) experiments have revealed that an in-plane uniaxial strain on {\BFA}  
induces an enhancement of the low-energy spin fluctuations along the $c$-axis 
in the paramagnetic state above $T_N$ \cite{Kissikov}.  However, it is unclear whether 
the applied uniaxial pressure can actually modify the 
collinear AF structure of the system [Fig. 1(a)] \cite{qhuang,kim2011}.

In this work, we use polarized neutron scattering and unpolarized neutron diffraction to 
demonstrate that an in-plane uniaxial pressure necessary to detwin {\BFA} also induces
a $c$-axis ordered magnetic moment and changes the easy axis of the collinear AF structure 
around the magnetic/nematic 
critical scattering temperature regime where the applied pressure has 
a large impact on the lattice structure of the system [Figs. 1(c-g)] \cite{Lu2016}. 
In addition, we find that the applied pressure induces 
$c$-axis polarized critical spin fluctuations that diverge near $T_N/T_s$, confirming 
the results of NMR experiments \cite{Kissikov}.  
Therefore, uniaxial pressure on {\BFA} that breaks the tetragonal lattice 
symmetry also induces changes in the magnetic easy axis
near the critical regime of the AF/nematic phase transitions, indicating that 
 the intrinsic electronic and magnetic properties of the system near $T_N/T_s$
are much different from naive expectations.

\section{Results}

\begin{figure}[t]
\includegraphics[width=7.4cm]{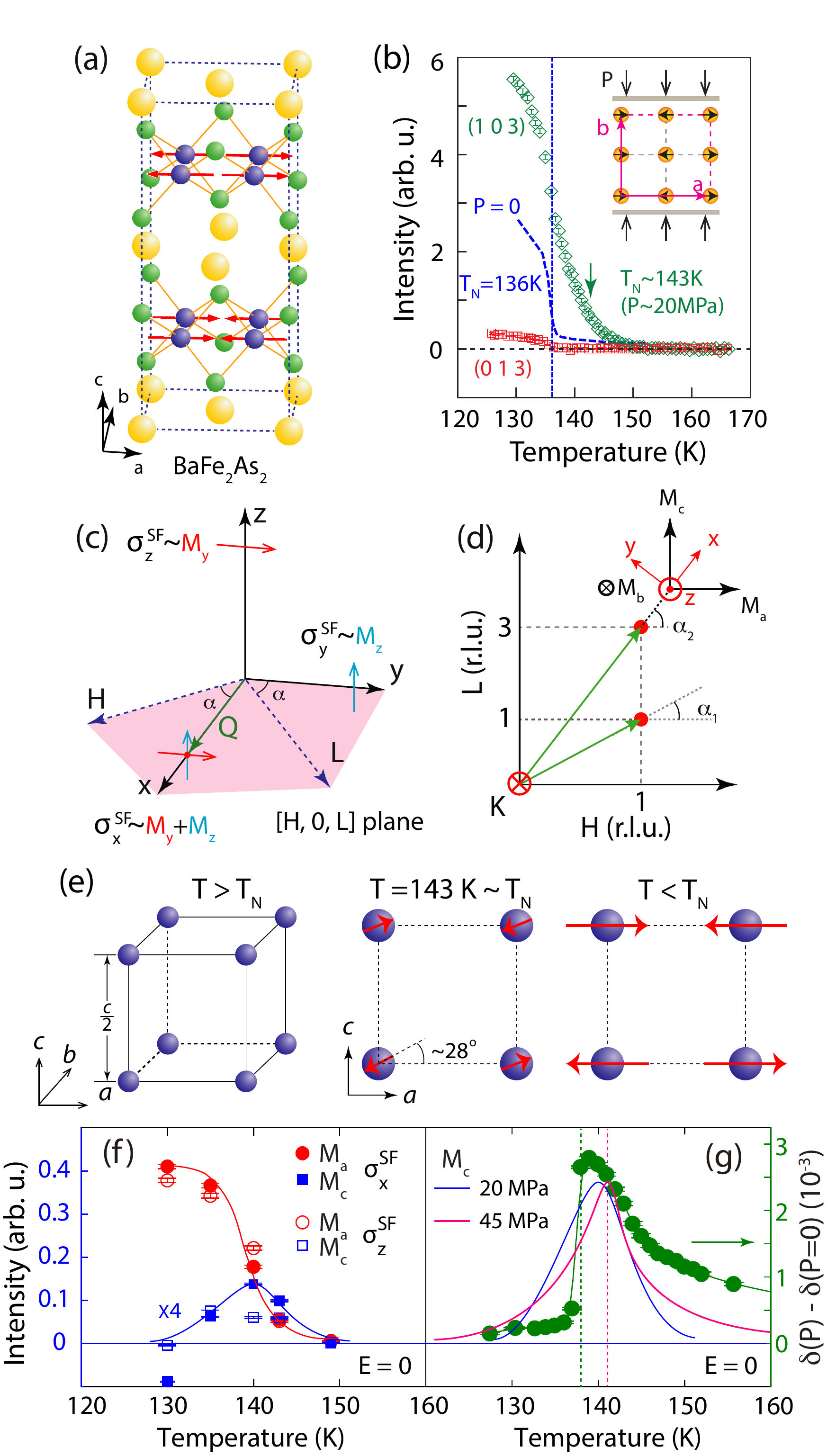}
\caption{{\bf Summary of the effect of uniaxial pressure
on crystalline lattice and magnetic structures of Ba$_2$Fe$_2$As$_2$.}  
(a) Crystal and AF structure of BaFe$_2$As$_2$.  
The red arrows indicate the $a$-axis direction of magnetic moments.  
(b) Magnetic order parameters measured at ${\bf Q}_1=(1, 0, 3)$ and $(0, 1, 3)$ 
under uniaxial pressure, revealing $T_N=143$ K. The blue dashed curve denotes the magnetic order parameter measured on a strain-free sample with $T_N=136$ K.
(c) Scattering geometry of polarized neutron scattering experiment in the $[H,0,L]$ plane. 
(d) The reciprocal space, where the fluctuating moments along the $a$-, $b$-, and $c$-axis directions are marked as $M_a$, $M_b$, and $M_c$, respectively.  (e) Spin arrangements of BaFe$_2$As$_2$ in the paramagnetic (left), near $T_N$ (middle), and low-temperature AF state. 
(f) Temperature dependence of the static ordered
magnetic moment $M_a$ and $M_c$ as determined from $\sigma_{x,z}^{SF}$ at $(1,0,1)$ and $(1,0,3)$.  The vertical error bars are estimated errors from fits-to-order parameters. (g) Comparison of temperature dependence of the strain-induced lattice distortion
from Ref. \cite{Lu2016} and our estimated $M_c$ at $\sim$20 (blue solid line) 
and $\sim$45 (pink solid line) MPa.   }
\label{fig1}
\end{figure}

{\bf Collinear magnetic order in twinned BaFe$_2$As$_2$.} Without external uniaxial pressure, {\BFA} exhibits separate weakly first-order magnetic
and second-order structural phase transitions ($T_s > T_N$ by $\sim$0.75 K) \cite{kim2011}. 
The spins within each FeAs layer are collinear and arranged antiferromagnetically along the 
$a$-axis and ferromagnetically along the $b$-axis 
of orthorhombic structure with lattice parameters of $a$ and $b$, respectively ($a>b$). Along the out-of-plane direction,
spins are arranged antiferromagnetically within one chemical unit cell 
(lattice parameter $c$), but have no net magnetic moment 
along the $c$-axis [Fig. 1(a)] \cite{qhuang,kim2011}. For a collinear Ising antiferromagnet with second order (or weakly first order) magnetic phase transition, 
magnetic critical scattering with moments polarized along the longitudinal (parallel to the ordered moment or $a$-axis) direction should diverge at $T_N$, while spin fluctuations with moments polarized transverse to the ordered moment ($b$- and $c$-axis) direction should not diverge \cite{collins,birgeneau70,birgeneau75,MnF2,tseng16}. Unpolarized \cite{Wilson10} and polarized  \cite{yuli} neutron scattering experiments on strain-free {\BFA} confirm 
this expectation.  While 
the longitudinal component ($M_a$) of the magnetic 
critical scattering, defined as low-energy spin fluctuations polarized along the $a$-axis direction,  
diverges at $T_N$, the transverse components $M_b$ and $M_c$ along the $b$ and $c$-axis, respectively [Figs. 1(c,d)], do not diverge at $T_N$.  

{\bf Effect of uniaxial pressure on lattice parameters of BaFe$_2$As$_2$.} The in-plane uniaxial pressure-induced tetragonal symmetry-breaking lattice distortion
[$\delta(P\neq 0)-\delta(P=0)$, where $\delta=(a-b)/(a+b)$] has 
a Curie-Weiss temperature dependence in the paramagnetic state and peaks near $T_N/T_s$,
but is greatly suppressed below $T_N/T_s$ when the intrinsic orthorhombic lattice of {\BFA} 
sets in [Fig. 1(g)] \cite{Lu2016}. In the paramagnetic state, NMR experiments on {\BFA} suggest that 
an in-plane uniaxial strain can induce 
a diverging $c$-axis polarized spin susceptibility $\chi^{\prime\prime}_{c}$, which equals to $M_c$ 
in the zero energy limit, on approaching $T_N/T_s$ \cite{Kissikov}. 
Since $c$-axis polarized low-energy spin fluctuations do not diverge around $T_N/T_s$ 
in the strain-free {\BFA} \cite{yuli}, it is important
to confirm the NMR results and determine if the diverging $\chi^{\prime\prime}_{c}$ is a precursor of
a new magnetic order with a component along the $c$-axis [Fig. 1(f)] \cite{collins}.

\begin{figure}[t]
\includegraphics[width=8cm]{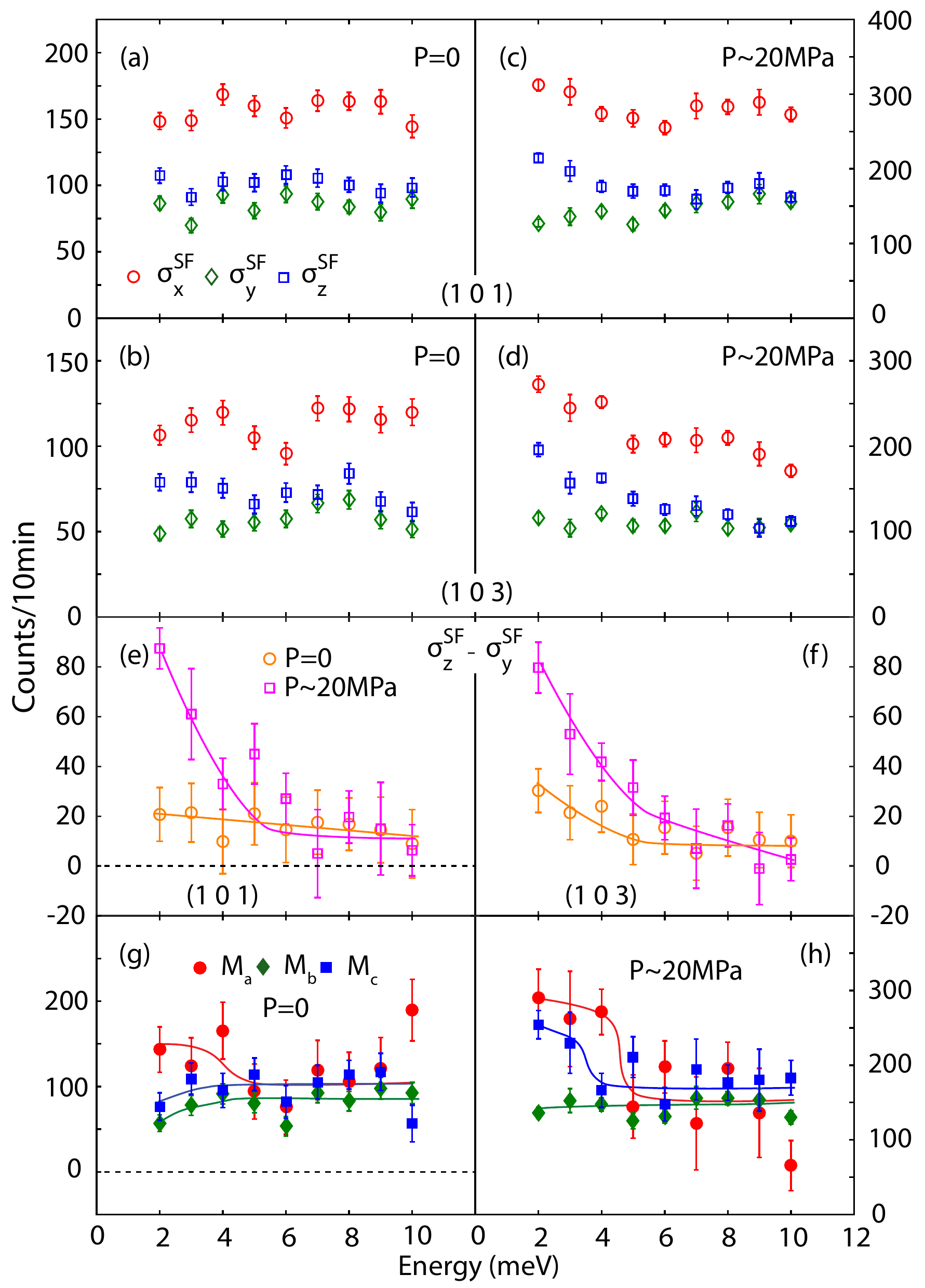}
\caption{{\bf The energy dependence of the neutron spin-flip magnetic scattering near
$T_N$ without and with uniaxial pressure.} 
(a-d) Energy scans of $\sigma_x^{SF}$ (red circle), $\sigma_y^{SF}$ (green diamond), and $\sigma_z^{SF}$ (blue square) under (a, b) $P=0$ and $T=138$ K \cite{yuli} and (c, d) $\sim$20 MPa and $T=145$ K at the two AF wave vectors $(1,0,1)$ and $(1,0,3)$. 
(e, f) Comparison of $P=0$ and $P\approx 20$ MPa $(\sigma_z^{SF}-\sigma_y^{SF})$ 
at $(1,0,1)$ and $(1,0,3)$. 
(g, h) Energy dependence of $M_a, M_b$, and $M_c$ extracted from the raw data in (a-d). 
The solid lines are guides to the eyes and the error bars represent 
one standard deviation.}
 \end{figure}

{\bf Neutron polarization analysis of spin excitations in detwinned BaFe$_2$As$_2$.}  
Our polarized neutron scattering experiments were carried out on the CEA 
CRG-IN22 triple-axis spectrometer equipped 
with Cryopad capability at the Institut Laue Langevin and the BT-7 triple-axis spectrometer at the NIST Center for Neutron Research.  The experimental setup for IN22 has been described in detail before \cite{yuli,Lipscombe,NQureshi2014,CWangPRX,YS1,CL1}, while polarized neutrons were controlled and 
analyzed using a polarized $^3$He filter on BT-7 \cite{lynn1,lynn2}. 
We have also carried out unpolarized
neutron diffraction experiments on BT-7 using an in-situ uniaxial 
pressure device \cite{Tam2017}.
The wave vector transfer ${\bf Q}$ in reciprocal space
in \AA$^{-1}$ is defined as ${\bf Q}= H {\bf a}^\ast+K {\bf b}^\ast+ L {\bf c}^\ast$, with 
${\bf a}^\ast=(2\pi/a){\bf \hat{a}}$, ${\bf b}^\ast=(2\pi/b){\bf \hat{b}}$, and  
${\bf c}^\ast=(2\pi/c){\bf \hat{c}}$, where $a\approx b\approx 5.6$ \AA, $c=12.96$ \AA, and $H$, $K$, $L$ are Miller indices. In this notation, the collinear AF structure of {\BFA} in Fig. 1(a) gives 
magnetic Bragg peaks at $[H,K,L]=[1,0,L]$ with $L=1,3,\dots$. The
magnetic responses of the system at a particular ${\bf Q}$ along the orthorhombic lattice $a$-, $b$-, and
$c$-axis directions are marked as $M_a$, $M_b$, and $M_c$, respectively [Figs. 1(a-d)].  The scattering plane is $[H,0,L]$. The incident neutrons are polarized along the ${\bf Q}$ ($x$), perpendicular to ${\bf Q}$
but in the scattering plane ($y$), and perpendicular to both ${\bf Q}$ 
and the scattering plane ($z$) [Fig. 1(c)].  In this geometry, 
the neutron spin-flip (SF) scattering cross sections $\sigma^{SF}_x$, $\sigma^{SF}_y$, and $\sigma^{SF}_z$
are related to the 
components $M_a$, $M_b$, and $M_c$ via $\sigma _x^{SF} =  
\frac{R}{R+1} M_y + \frac{R}{R+1} M_z +B$, 
$\sigma _y^{SF} = \frac{1}{R+1} M_y + \frac{R}{R+1} M_z +B$, and 
$\sigma _z^{SF} = \frac{R}{R+1} M_y + \frac{1}{R+1} M_z +B$, where $R$ is the flipping ratio 
($R=\sigma _{Bragg}^{NSF}/\sigma _{Bragg}^{SF}\approx 13$), 
$B$ is the background scattering, 
$M_y= \sin ^2 \alpha M_a + \cos ^2 \alpha M_c$ with $\alpha$ being the angle between
$[H,0,0]$ and ${\bf Q}$, and $M_z=M_b$ [Fig. 1(d)] \cite{yuli,Lipscombe,NQureshi2014,CWangPRX,YS1,CL1}.

Figure 1(b) compares the temperature dependencies of the $(1,0,3)$ magnetic Bragg 
peak for strain-free and strained {\BFA}.
At zero external pressure ($P=0$ and strain-free), the magnetic scattering shows an order parameter like increase below $T_N=136$ K \cite{yuli}.
When an uniaxial pressure of $P\approx 20$ MPa is applied along the $b$-axis of {\BFA}, 
the N$\rm \acute{e}$el temperature of the sample increases to $T_N=143$ K \cite{Tam2017}. The vanishingly
small magnetic scattering intensity at ${\bf Q}=(0,1,3)$ suggests that the sample is essentially $\sim$100\% detwinned [Fig. 1(b)].

\begin{figure}[t]
\includegraphics[scale=.45]{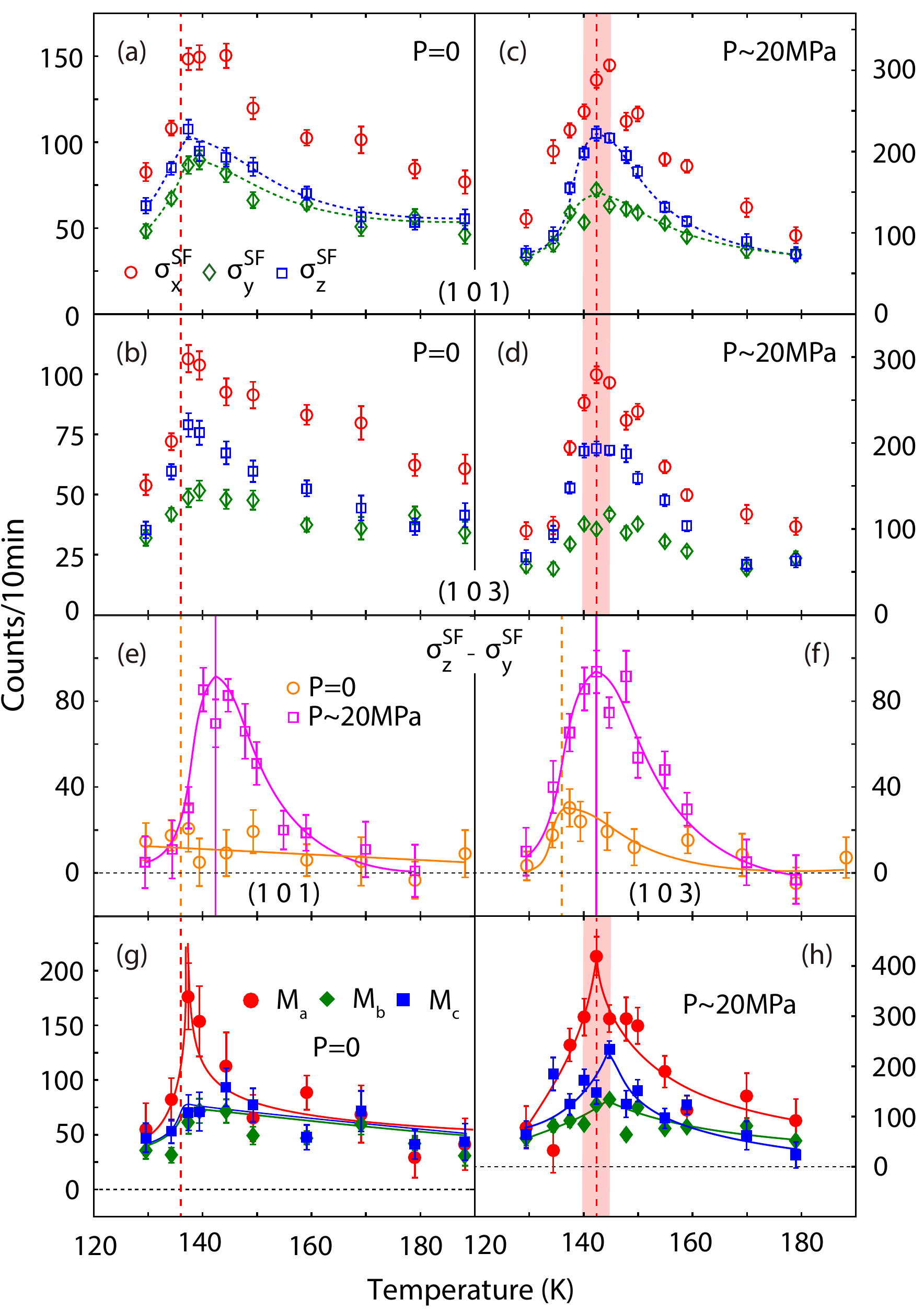}
\caption{{\bf Temperature dependence of the magnetic scattering across
$T_N$ at $E=2$ meV without and with uniaxial pressure.} 
Temperature dependence of $\sigma^{SF}_x$, $\sigma^{SF}_y$, and $\sigma^{SF}_z$ at $E=2$ meV of (a, b) uniaxial pressure-free \cite{yuli} and 
(c, d) pressured ($P\approx 20$ MPa) {\BFA} 
at (a, c) $(1,0,1)$ and (b, d) $(1,0,3)$.  
(e, f) Comparison of $P=0$ and $P\approx 20$ MPa $(\sigma_z^{SF}-\sigma_y^{SF})$ 
at $(1,0,1)$ and $(1,0,3)$. (g, h)
Temperature dependence of $M_a$, $M_b$, and $M_c$ at $E=2$ meV for (g) uniaxial pressure-free 
and (h) pressured sample estimated from the data in (a-d). 
The dotted and solid lines are guides to the eye and the error bars represent 
one standard deviation.
The vertical dashed and solid lines mark $T_N/T_s$ at $P=0$ and $P\approx 20$ MPa, respectively.}
\end{figure}

\begin{figure}[t]
\includegraphics[scale=.45]{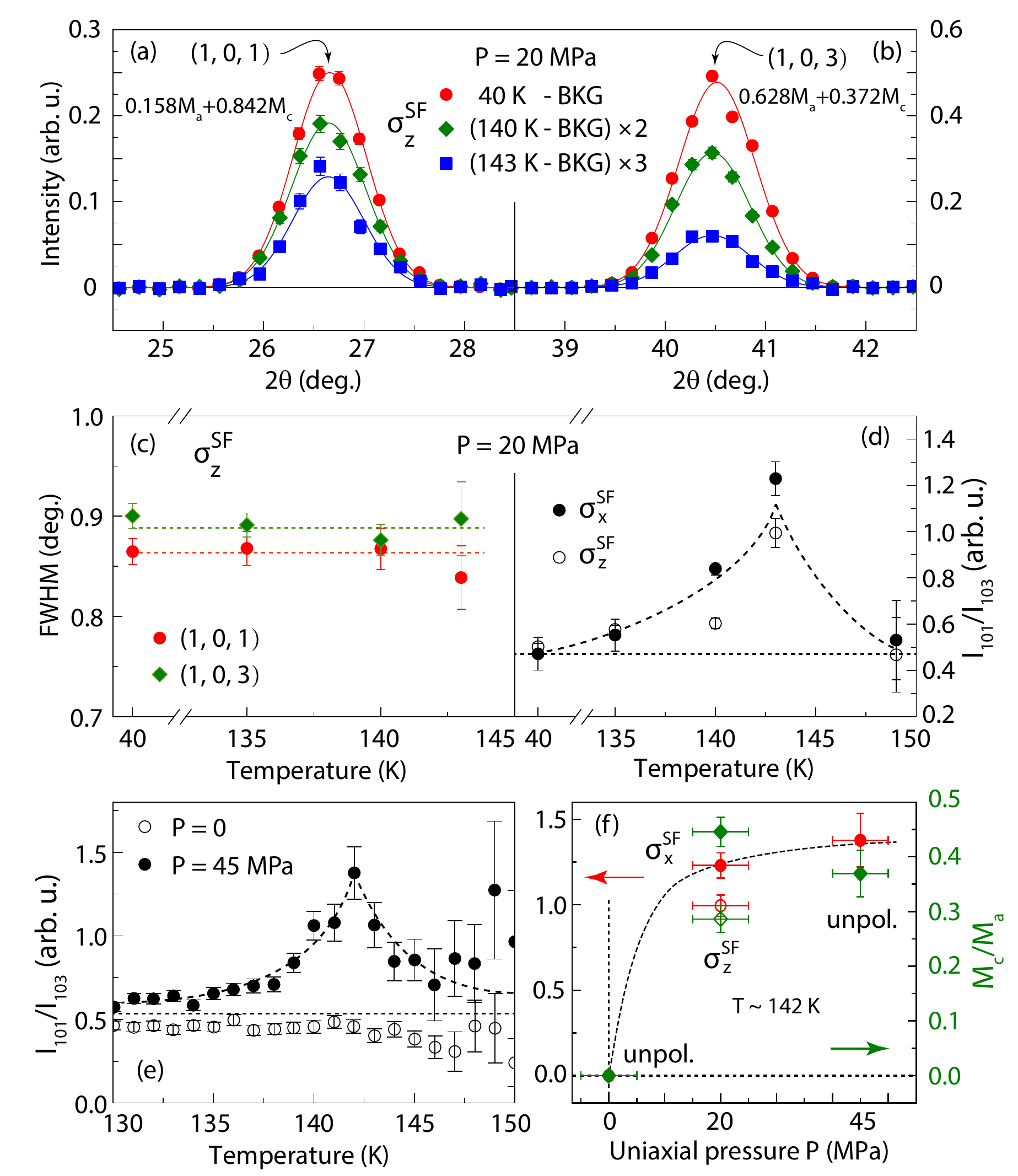}
\caption{{\bf Uniaxial pressure dependence of the magnetic order and correlations.}
Elastic $\theta/2\theta$ scans of $\sigma^{SF}_z$ 
across (a) $(1,0,1)$ and (b) $(1,0,3)$ at different temperatures and $P=20$ MPa. 
The data are collected on BT-7 using final neutron energy of 14.7 meV with instrumental
energy resolution of about $1.3$ meV. Similar scans of 
$\sigma^{SF}_z$ are discussed in \cite{SI}. $\sigma^{SF}_y$ was unavailable at the time of this experiment.
 (c) FWHM of the elastic 
$(1,0,1)$ and $(1,0,3)$ scans across $T_N/T_s$. (d) Temperature dependence of 
$I_{101}/I_{103} \propto (0.16 M_a + 0.84 M_c)/(0.63 M_a + 0.37 M_c)$. 
(e) Temperature dependence of $I_{101}/I_{103}$ at $P\approx 0$ and $\sim$45 MPa uniaxial
pressure obtained using in-situ uniaxial pressure device with unpolarized neutrons on BT-7. (f) Pressure dependence of $I_{101}/I_{103}$ (red symbols, left axis) and $M_c/M_a$ (green symbols, right axis) at $T \sim 142$ K. The data points for $P=0$ and 45 MPa were measured with unpolarized neutron scattering. The data points for $P=20$ Mpa were measured with polarized mode with the open (solid) symbols obtained from $\sigma^{SF}_z$ ($\sigma^{SF}_x$). The black dashed curve is a guide to the eye for the data points. “unpol.” denotes unpolarized neutron scattering measurements.
The vertical error bars in (a,b,d,e) represent statistical errors of
1 standard deviation. The error bars in (c) are estimated errors from fits to
magnetic Bragg peak widths.  The error bars in (f) are our estimated errors from
fits to magnetic order parameters and applied uniaxial pressure. 
  }
\end{figure}

Figures 2(a) and 2(b) show the energy dependence of 
$\sigma _x^{SF}$, $\sigma^{SF}_y$, and $\sigma^{SF}_z$ in the zero pressure 
paramagnetic state at $T\approx 1.015T_N\approx 138$ K for magnetic positions  
$(1,0,1)$ and $(1,0,3)$ \cite{yuli}.  Figures 2(c) and 2(d) show identical scans 
as those of Figs. 2(a) and 2(b), respectively, in the paramagnetic
state at $T\approx 1.014T_N\approx 145$ K with uniaxial pressure of 
$P\approx 20$ MPa.  Comparison of the Figs. 2(a) and 2(c) reveals  
that $\sigma^{SF}_z$ is clearly larger than $\sigma^{SF}_y$ below $\sim$5 meV 
in the uniaxial strained sample. Since the $(1,0,1)$ peak corresponds to $\alpha_1=23.4^\circ$ 
giving $M_y\approx 0.16 M_a + 0.84 M_c$ [Fig. 1(d)] \cite{yuli}, 
the increased $\sigma^{SF}_z$ in strained  
{\BFA} is mostly due to the increased $M_c$.  For the $(1,0,3)$ peak, which corresponds to 
$\alpha_2=52.4^\circ$, $M_y\approx 0.63 M_a + 0.37 M_c$, and the scattering is therefore much less sensitive to strain-induced changes in $M_c$.  
To conclusively determine the effect of uniaxial pressure on $M_c$, 
we consider $\sigma _z^{SF}-\sigma _y^{SF}\propto
M_y-M_b $. Since $M_b$ (or $\sigma _y^{SF}$) does not diverge at $T_N$ or 
change as a function of uniaxial pressure as seen in NMR \cite{Kissikov} and neutron polarization analysis [Figs. 2(a-d)], the effect of uniaxial pressure can be seen directly from the energy
dependence of  $\sigma _z^{SF}-\sigma _y^{SF}$ at the $(1,0,1)$ [Fig. 2(e)] and 
$(1,0,3)$ [Fig. 2(f)].  Without uniaxial pressure, $\sigma _z^{SF}-\sigma _y^{SF}$ 
 does not diverge at the $(1,0,1)$ position but diverges at $(1,0,3)$
at low energies consistent with the expectation that spin fluctuations at $(1,0,1)$ is mostly probing $M_c$.  With uniaxial pressure, we see clear divergence of low-energy spin fluctuations at 
 $(1,0,1)$ below $\sim$5 meV, thus unambiguously confirming 
the uniaxial pressure induced $M_c$ enhancement around $T_N$ observed in 
NMR experiments \cite{Kissikov}.  
To further clarify the energy dependence of $M_a$, $M_b$, and $M_c$,
we estimate these components from measurements at the 
$(1,0,1)$ and $(1,0,3)$ positions as described in Ref. \cite{yuli}.
By comparing the energy dependence of $M_a$, $M_b$, and $M_c$ in strain-free [Fig. 2(g)] and strained [Fig. 2(h)] {\BFA}, we again see that the effect of uniaxial strain is to enhance $M_c$ below
about 4 meV, consistent with the NMR measurements which probe $M_c$ or $\chi^{\prime\prime}_{c}$ in the zero energy limit \cite{Kissikov}.

To demonstrate further the effect of uniaxial strain on the magnetic critical scattering of {\BFA}, we show in Fig. 3 the temperature dependence of $\sigma^{SF}_x$, $\sigma^{SF}_y$, and $\sigma^{SF}_z$ at $E=2$ meV for the strain-free [Figs. 3(a,b)] \cite{yuli} and strained [Figs. 3(c,d)] samples. 
At ${\bf Q}=(1,0,1)$, uniaxial strain clearly enhances  $\sigma^{SF}_z$
around $T_N/T_s$, where $\sigma^{SF}_z\approx M_y\approx 0.16 M_a + 0.84 M_c$, again 
consistent with the strain enhanced $\chi^{\prime\prime}_{c}$ in the NMR measurements \cite{Kissikov}. 
Figures 3(b) and 3(d) show similar measurements at ${\bf Q}=(1,0,3)$, which reveal much
less enhancement of $\sigma^{SF}_z$ because 
$\sigma^{SF}_z\approx M_y\approx 0.63 M_a + 0.37  M_c$.
Figure 3(e) shows temperature dependence of 
$\sigma _z^{SF}-\sigma _y^{SF}$ across $T_N$ at the $(1,0,1)$ 
peak without and with uniaxial pressure.  Since spin fluctuations at the $(1,0,1)$ position
is mostly sensitive to $M_c$, we see no divergence across $T_N$ in zero pressure case.  Upon
application of a $\sim$20 MPa uniaxial pressure, the scattering clearly reveals a digerving 
behavior at the pressured enhanced $T_N$ (solid vertical line) [Fig. 3(e)].  Similar measurements
at the $(1,0,3)$ position, which is more sensitive to $M_a$, show diverging magnetic
scattering at $T_N$ with and without uniaxial pressure consistent with the NMR 
results [Fig. 3(f)] \cite{Kissikov}.  Figures 3(g) and 3(h) show the temperature dependencies of the estimated  
$M_a$, $M_b$, and $M_c$ for strain-free and strained {\BFA}, respectively, using the data in Figs. 3(a-d).  Comparing with the normal behavior of the strain-free {\BFA} [Fig. 3(g)],
the $M_c$ in strained {\BFA} clearly diverges around $T_N/T_s$ [Fig. 3(h)], although the error bars of the data became worst after the data manipulation \cite{SI}.

{\bf Effect of uniaxial pressure on static AF order of BaFe$_2$As$_2$.} In principle, a diverging dynamic spin susceptibility in the paramagnetic state of a system is an indication of the eventual magnetic order 
below $T_N$ \cite{collins,birgeneau70,birgeneau75,MnF2,tseng16}. 
For strain-free {\BFA}, the magnetic ordered moment is along the $a$-axis with no net moment along the $b$-axis and $c$-axis directions \cite{qhuang,kim2011}. Therefore, 
only the $M_a$ component of the
spin susceptibility diverges at $T_N$ [Figs. 3(e,f,g)] \cite{yuli}. 
The observation of a diverging $M_c$ in strained {\BFA},
in addition to the usual diverging $M_a$ [Figs. 3(e,f,h)], suggests that the applied strain may induce static magnetic ordered moment along the $c$-axis. To test this hypothesis, we carried out polarized neutron diffraction measurements on {\BFA} as a function of uniaxial pressure, focusing on the temperature and neutron polarization dependence 
of the scattering at ${\bf Q}=(1,0,L)$ with $L=0,1,2,3,$ and 5.  At wave vectors 
$(1,0,0)$ and $(1,0,2)$, there is no evidence of magnetic scattering, consistent with 
uniaxial pressure-free {\BFA} \cite{SI}. 

Figures 4(a) and 4(b) show $\theta/2\theta$ scans of 
$\sigma^{SF}_z$ around $(1,0,1)$ and $(1,0,3)$, respectively, at different temperatures. 
Since $\sigma^{SF}_z$ at these two wave vectors probes different combinations of $M_a$ and $M_c$,
one can obtain magnitudes of the static ordered moments along the $a$-axis and $c$-axis 
directions at these temperatures. 
Figure 4(c) shows the temperature dependencies of the
full-width-at-half-maximum (FWHM) of these peaks, indicating that 
the spin-spin correlation lengths are instrumental resolution limited and temperature independent.
Figure 4(d) plots the magnetic scattering intensity ratio between $(1,0,1)$ ($I_{101}$) 
and $(1,0,3)$ ($I_{103}$), which measures the relative strength of $M_c$ and reveals a clear peak around $T_N/T_s$. 

To further determine the effect of uniaxial pressure on $c$-axis ordered moment and its pressure dependence, we carried out unpolarized neutron diffraction measurements focusing
on the magnetic scattering intensity ratio between $(1,0,1)$ ($I_{101}$) 
and $(1,0,3)$ ($I_{103}$) using an in-situ uniaxial pressure device.  Since our polarized
neutron diffraction measurements revealed no ordered moment $M_b$, we used unpolarized
neutron diffraction on BT-7 to improve the statistics of the data across $T_N$. 
Figure 4(e) compares the measured $I_{101}/I_{103}$ from 130 K to 150 K 
at $P\approx 0$ and 45 MPa uniaxial pressure. Consistent with earlier work \cite{qhuang,kim2011}, $I_{101}/I_{103}$ is approximately temperature independent
across $T_N$ at $P\approx 0$, thus indicating that the 
internal strain of the system does not induce a $c$-axis ordered moment.   
Upon applying an uniaxial pressure of $P\approx 45$ MPa, the identical
measurement shows a dramatic peak at $T_N$, thus
confirming the results of Figs. 4(a-d).  Figure 4(f) shows 
the uniaxial pressure dependence of the measured $M_c/M_a$ at $T_N$, suggesting
that the ordered $c$-axis moment saturates with increasing pressure above 45 MPa.

Figure 1(f) shows the temperature dependencies of the magnetically ordered moments along the $a$-axis ($M_a$) and $c$-axis ($M_c$) directions obtained 
by comparing $\sigma^{SF}_x$ and $\sigma^{SF}_z$ at wave vectors $(1,0,1)$ and $(1,0,3)$ \cite{SI}.    
In the low temperature AF ordered state, the strain-free and 
strained {\BFA} have the standard
collinear AF structure with no evidence of $M_c$ 
[right panel in Fig. 1(e) and Fig. 4(e)] \cite{qhuang,kim2011}.  
On warming to 143 K just below $T_N$, the easy-axis tilts from the $a$-axis towards the $c$-axis 
with an angle of $\sim$28$^\circ$ [middle panel in Fig. 1(e)].  Finally, on warming to temperatures well above $T_N$, there is no static AF order [left panel in Fig. 1(e)].
Figure 1(g) shows the temperature dependence of $M_c$ at $\sim$20 (blue solid line) 
and $\sim$45 (pink solid line) MPa uniaxial
pressure, compared with the uniaxial strain-induced lattice
distortion $\delta(P\approx 20\ {\rm MPa})-\delta(P=0)$ (green solid circles and lines)  obtained from neutron Larmor diffraction
experiments \cite{Lu2016}.  The similarity of the data suggests 
that the $c$-axis aligned magnetic moment arises 
from the uniaxial pressure-induced lattice distortion.

\section{Discussion}

Theoretically, the in-plane electronic anisotropy of the iron pnictides is expected to couple linearly to the lattice orthorhombicity by the Ginzburg-Landau
free-energy formalism if one ignores the effect of spin-orbit coupling induced magnetic
anisotropy \cite{AEBohmerCRP,Lu2016}.  From this perspective, in-plane uniaxial strain should only induce in-plane electronic anisotropy.  The discovery of a $c$-axis ordered magnetic moment coupled exclusively with uniaxial pressure-induced lattice distortion suggests that such an effect
cannot be only associated with the lattice orthorhombicity of the system, as $M_c$ becomes vanishingly small in the low-temperature orthorhombic phase with 
large in-plane lattice distortion. This is also different from the $c$-axis moment 
AF structure in Ba$_{1-x}$K$_x$Fe$_2$As$_2$ in the sense that the $c$-axis order appears exclusively 
in the tetragonal phase \cite{Avci2014,Allred2016}, while the $c$-axis moment appears in {\BFA} 
only near the peak of the nematic 
susceptibility around $T_N/T_s$. Although there is currently no theoretical
understanding of this observation, it must arise from 
spin-orbit coupling induced magnetic anisotropy \cite{Scherer}. 
Our discovery opens a new avenue to control magnetic order in nematic materials using mechanical strain instead of magnetic fields. The strong coupling of the $c$-axis aligned magnetic order with an in-plane pressure-induced lattice distortion offers
the potential for the next generation of mechanical-strain-controlled magnetic switches. One must consider the presence of the magnetically ordered moment along the $c$-axis in mechanically detwinned iron pnictides in order to understand their intrinsic electronic, magnetic, and nematic properties. 

Alternatively, our observations are also consistent with strain inducing a proximate $XY$ spin anisotropy near $T_N$/$T_s$. In this scenario, while $a$-axis is energetically favorable in terms of spin anisotropy, $c$-axis is very close. This allows for a distribution of large (resolution-limited but not long-ranged ordered) and long-lived (quasi-static) collinear magnetic domains, with their collinear spin direction in the $ac$-plane. The ratio between $I_{101}/I_{103}$ [Figs. 4(d,e)] is then a measure of the distribution, reflective of the difference in spin anisotropy energies along the $a$- and $c$-axis. Similar to when the easy axis tilts from $a$-axis towards $c$-axis under strain [Fig. 1(e)], the change to a proximate $XY$ spin anisotropy under strain also indicates of a large and highly unusual effect of strain on the spin anisotropy.

In conclusion, we have used polarized and unpolarized neutron scattering to study the magnetic structure and critical scattering in uniaxial strained {\BFA}. We find that the uniaxial pressure necessary to make single domain samples of {\BFA} also induces $c$-axis polarized critical magnetic scattering and static magnetic order around $T_N/T_s$. The size of the $c$-axis ordered moment is associated with the uniaxial pressure-induced lattice distortion, instead of the lattice orthorhombicity. These results indicate that in addition to detwinning {\BFA}, uniaxial pressure applied on the sample actually modifies the magnetic structure of the system. Therefore, infrared \cite{Nakajima}, angle resolved photoemission \cite{MYi2017}, and Raman spectroscopy \cite{Ren2015,Baum} experiments on mechanically detwinned {\BFA} near the 
magnetic and nematic phases should be re-examined to take into 
account the effect of strain-induced change to the spin anisotropy on the in-plane electronic and magnetic properties. 

\section{Methods}

{\bf Sample preparation and experimental details.} BaFe$_2$As$_2$ single crystals
were grown by the self-flux method using the same growth procedure as described before \cite{Lu14}.
Our polarized inelastic neutron scattering experiments were carried out using
the IN22 CEA-CRG triple-axis spectrometer at the Institut 
Laue-Langevin, Grenoble, France \cite{ILL}. Polarized neutrons were produced using a focusing Heusler monochromator
and analyzed with a focusing Heusler analyzer with a final
wave vector of $k_f=2.662$ \AA$^{-1}$.  The experimental setups for uniaxial pressured
and pressure freed experiments are identical. However, it is difficult to directly compare
the scattering intensity of these two experiments since the sample masses, their relative positions in the beam, and background scattering of these two experiments are different.  
Nevertheless, one can safely compare the relative intensity changes of these two experiments.
The polarized elastic neutron scattering experiments were carried out on BT-7 
utilizing $^3$He polarizers immediately before and after the sample 
at NIST center for neutron research, Gaithersburg, Maryland, USA \cite{lynn1,lynn2}. 
The unpolarized neutron diffraction experiments in Fig. 4(e) were carried out 
using a pyrolytic graphite monochromator and analyzer with pyrolytic graphite filter in the beam. 
Experiments on twinned BaFe$_2$As$_2$ without external
uniaxial pressure were performed on $\sim$12-g aligned single crystals 
as described before \cite{yuli}.  The polarized inelastic neutron scattering experiments 
on uniaxial pressured detwinned BaFe$_2$As$_2$ were performed using 12 pieces cut    
 single crystals ($\sim$3-g, Fig. S1) \cite{xylu18}. 
The BT-7 measurements were carried out on a single piece of 
BaFe$_2$As$_2$ mounted on a newly built in-situ uniaxial pressure device and the
neutron wave vectors are set at  $k_i=k_f=2.662$ \AA$^{-1}$.

{\bf Determination of ${\bf M_a}$, ${\bf M_b}$ and ${\bf M_c}$.} 
In our previous polarized neutron scattering studies of iron pnictides, we have established the method for determining the spin-fluctuation components $M_{\beta}$ ($\beta=a, b, c$) along the lattice axes via comparing the spin-flip scattering $\sigma^{SF}_{\gamma}$ ($\gamma = x, y, z$) at two equivalent magnetic wave vectors (such as $\mathbf{Q_1}=(1, 0, 1)$ and $\mathbf{Q_2}=(1, 0, 3)$ as shown in Fig. 1 of the main text).  The definition of the directions $x, y$ and $z$ are described in Fig. 1. $\sigma^{SF}_{\gamma}$ is directly related to the spin-fluctuation components by:

\begin{equation}
\left\{
\begin{array}{cccccc}
\sigma_{x}^{\rm SF}({\bf Q})=F^2({\bf Q})\sin^2\alpha_{\bf Q}\frac{R}{R+1}M_{a}+F^2({\bf Q})\frac{R}{R+1}M_{b}+F^2({\bf Q})\cos^2\alpha_{\bf Q}\frac{R}{R+1}M_{c}+B({\bf Q}),\\
\\[1pt]
\sigma_{y}^{\rm SF}({\bf Q})=F^2({\bf Q})\sin^2\alpha_{\bf Q}\frac{1}{R+1}M_{a}+F^2({\bf Q})\frac{R}{R+1}M_{b}+F^2({\bf Q})\cos^2\alpha_{\bf Q}\frac{1}{R+1}M_{c}+B({\bf Q}),\\
\\[1pt]
\sigma_{z}^{\rm SF}({\bf Q})=F^2({\bf Q})\sin^2\alpha_{\bf Q}\frac{R}{R+1}M_{a}+F^2({\bf Q})\frac{1}{R+1}M_{b}+F^2({\bf Q})\cos^2\alpha_{\bf Q}\frac{R}{R+1}M_{c}+B({\bf Q})\\
\end{array}
\right.
\end{equation}
where $\alpha$ is the angle between (1, 0, 0) and {\bf Q} (Fig. 1), $F({\bf Q})$ is magnetic form factor of Fe$^{2+}$, $R$ is the flipping ratio 
($R=\sigma _{Bragg}^{NSF}/\sigma _{Bragg}^{SF}\approx 13$), 
and $B$ is the polarization-independent background scattering.

From Eq. (1), we can get four equations for our results on $\mathbf{Q_1}$ and $\mathbf{Q_2}$:
\begin{equation}
\left\{
\begin{array}{cccccc}
\sigma_{x}^{\rm SF}({\bf Q_1})-\sigma_{y}^{\rm SF}({\bf Q_1})=\frac{R-1}{R+1}F^2({\bf Q_1})[\sin^2\alpha_1M_{a}+\cos^2\alpha_1M_{c}],\\
\\[1pt]
\sigma_{x}^{\rm SF}({\bf Q_2})-\sigma_{y}^{\rm SF}({\bf Q_2})=r\frac{R-1}{R+1}F^2({\bf Q_2})[\sin^2\alpha_2M_{a}+\cos^2\alpha_2M_{c}],\\
\\[1pt]
\sigma_{x}^{\rm SF}({\bf Q_1})-\sigma_{z}^{\rm SF}({\bf Q_1})=\frac{R-1}{R+1}F^2({\bf Q_1})M_{b},\\
\\[1pt]
\sigma_{x}^{\rm SF}({\bf Q_2})-\sigma_{z}^{\rm SF}({\bf Q_2})=r\frac{R-1}{R+1}F^2({\bf Q_2})M_{b},\\
\end{array}
\right.
\end{equation}
in which $r$ is the intensity ratio factor between ${\bf Q_1}$ and ${\bf Q_2}$ to account for the differences in sample illumination volume and the convolution with instrumental resolution. The third and fourth equations in Eq. (2) can be used to determine the ratio $r$ and $M_b$, and the first two equations for $M_a$ and $M_c$. More details concerning the determination of the spin-fluctuation components $M_a$, $M_b$ and $M_c$ can be find elsewhere \cite{CL1}.  Although this method can determine the values of ${M_a}$, ${M_b}$ and ${M_c}$, it also results in large error bars of their values.  To more accurately determine the effect of uniaxial pressure on $M_a$ and $M_c$, we consider the differences between  
$\sigma_{z}^{\rm SF}({\bf Q})-\sigma_{y}^{\rm SF}({\bf Q})$ at ${\bf Q_1}$ and ${\bf Q_2}$.

\begin{equation}
\left\{
\begin{array}{cccccc}
\sigma_{z}^{\rm SF}({\bf Q_1})-\sigma_{y}^{\rm SF}({\bf Q_1})=\frac{R-1}{R+1}F^2({\bf Q_1})[\sin^2\alpha_1M_{a}+\cos^2\alpha_1M_{c}-M_{b}]\propto 0.16M_a+0.84M_c-M_b,\\
\\[1pt]
\sigma_{z}^{\rm SF}({\bf Q_2})-\sigma_{y}^{\rm SF}({\bf Q_2})=r\frac{R-1}{R+1}F^2({\bf Q_2})[\sin^2\alpha_2M_{a}+\cos^2\alpha_2M_{c}-M_{b}]\propto 0.63M_a+0.37M_c-M_b,\\
\end{array}
\right.
\end{equation}

As $M_b$ does not diverge in uniaxial pressured and pressure-free cases \cite{Kissikov}, a comparison 
of $\sigma_{z}^{\rm SF}({\bf Q_1})-\sigma_{y}^{\rm SF}({\bf Q_1})$ raw data should be most sensitive 
to changes in $M_c$, while $\sigma_{z}^{\rm SF}({\bf Q_2})-\sigma_{y}^{\rm SF}({\bf Q_2})$ should be sensitive
to changes in both $M_a$ and $M_c$.  The outcome of this analysis is shown in Figs. 2(e,f), 3(e,f). 

In our polarized neutron diffraction experiment at BT-7, we have only measured $\sigma^{SF}_x$ and $\sigma^{SF}_z$. In elastic channel, $M_{\beta}$ is proportional to the square of the ordered moment ($m_{\beta}$). The determination of $M_{\beta}$ follows the same method as described in Eqs. (1) and (2). But we need to apply the Lorentz factor ($L=\frac{1}{\sin2\theta}$) as we use the integrated intensity of $\theta-2\theta$ scan to calculate $M_{\beta}$ \cite{tas}, where $2\theta$ is the scattering angle for {\bf Q}.  Moreover, since no divergence of critical spin fluctuations were observed along the $b$ axis, we can assume the absence of static ordered moment ($M_b=0$) (even if we consider that quasi-elastic spin fluctuations along $b$ axis within the energy resolution of the elastic scattering could be included in $\sigma^{SF}_x$ and $\sigma^{SF}_z$, it can be neglected at least in $\sigma^{SF}_z$ because of the small pre-factor $\frac{1}{R+1}\approx0.07$ before $M_b$).
Then Eq. (2) can be written as:

\begin{equation}
\left\{
\begin{array}{cccccc}
\sigma_{x}^{\rm SF}({\bf Q_1})=\sigma_{z}^{\rm SF}({\bf Q_1})=\frac{1}{\sin2\theta_1}\frac{R}{R+1}F^2({\bf Q_1})[\sin^2\alpha_1M_{a}+\cos^2\alpha_1M_{c}],\\
\\[1pt]
\sigma_{x}^{\rm SF}({\bf Q_2})=\sigma_{z}^{\rm SF}({\bf Q_2})=r\frac{1}{\sin2\theta_2}\frac{R}{R+1}F^2({\bf Q_2})[\sin^2\alpha_2M_{a}+\cos^2\alpha_2M_{c}],\\
\end{array}
\right.
\end{equation}

Given the magnetic moment is polarized along $a$ axis at $40$K$<<T_N$ with $m_a\approx0.87\ \mu_B$, we can get $r$, solve $M_a$ and $M_c$ from both $\sigma^{SF}_x$ and $\sigma^{SF}_z$, and determine the magnitude of the $c$-axis moment induced by uniaxial strain. Taking $m_a=0.87\mu_B$ at $40$ K, we can get $m_a$ and $m_c$ at other temperatures using the data points shown in Fig. 1(f). From $\sigma^{SF}_z$, we get $m_a\approx0.23\pm0.05\ \mu_B$ and $m_c\approx0.12\pm0.03\ \mu_B$ at 143K, resulting in a canting angle of $\sim 28^{\circ}$ at this critical temperature. The calculated canting angles are estimated to be about $14^{\circ}$ at 140 K and 149 K, and gradually decreases to zero below 135K.

{\bf $\sigma^{SF}_{x, y, z}$ and $M_{a, b, c}$ below and well above $T_N$ 
at the AF ordering wave vectors.}  
Fig. S2 shows the results of $\sigma^{SF}_{\gamma}$ ($\gamma=x, y, z$) below and well above $T_N$ under zero and  $P \sim 20$ MPa.  At $T=135$ K ($<T_N$),
 $\sigma^{SF}_{\gamma}$'s for uniaxial pressure-free and pressured cases are shown in Figs. S2(a-d).
A comparison of $\sigma_{z}^{\rm SF}({\bf Q_1})-\sigma_{y}^{\rm SF}({\bf Q_1})$ scattering 
at $P=0$ and $\sim$20 MPa in Fig. S2(e) suggests that the applied uniaxial pressure
may enhance $M_c$ around $\sim$8 meV.  Similar data at ${\bf Q_2}$ in Fig. S2(f)
suggest that the effect of uniaxial pressure is limited on $M_a$ at this temperature.  
Figs. S2(g,h) shows as the converted $M_a, M_b$ and $M_c$ at $T=135$ K. 
At $T<T_N$, the data with $P\sim20$ MPa is qualitatively consistent with that measured on the $P=0$ sample, except that both the $M_a$ and $M_b$ are gapped below $E>10$ meV and $\sim 6$ meV, respectively, while only $M_a$ is gapped below $6$ meV for the $P=0$ sample. Note $T_N$ is $\sim 136$K for $P=0$ and $\sim 143$ K for $P\sim 20$ MPa. 
In relative temperature $T/T_N$, 135K  is much lower in the 
$P\sim 20$ MPa sample ($0.94T_N$) than that in free-standing sample ($0.99T_N$),  
thus the spin fluctuations are further gapped. For temperatures well above $T_N$ [Figs. S2(i-p)], spin-flip scattering becomes very weak and no qualitative difference were observed for $P=0$ and $P\sim 20$ MPa.

{\bf Comparison of ${\bf M_{\beta}}$ at {\bf Q}=(1, 0) and (0, 1).} To determine if the uniaxial pressure induced
$M_c$ at the AF wave vector ${\bf Q}=(1, 0)$ is compensated by magnetic scattering reduction at $(0,1)$, we compare
 $\sigma^{SF}_{\gamma}$ between ${\bf Q} = (1, 0, L)$ and $(0, 1, L)\ (L=1, 3)$ at $T=145$ K [Fig. S3(a-d)]. 
Figures S3(e) and (f) show the energy dependence of 
$\sigma_{z}^{\rm SF}({\bf Q})-\sigma_{y}^{\rm SF}({\bf Q})$ 
at ${\bf Q}=(1,0,1)/(0,1,1)$ and ${\bf Q}=(1,0,3)/(0,1,3)$, respectively.  
Compared with clear magnetic intensity gains below $\sim$6 meV at 
 the AF wave vectors ${\bf Q_1}=(1,0,1)$ and ${\bf Q_2}=(1,0,3)$, paramagnetic
scattering at ${\bf Q}=(0,1,1)$ and $(0,1,3)$ is isotropic in spin space as illustrated by the zero values 
of $\sigma_{z}^{\rm SF}({\bf Q})-\sigma_{y}^{\rm SF}({\bf Q})$ at these wave vectors. 
Figures S3(g) and (h) show the energy dependence of $M_a$, $M_b$, and $M_c$ 
extracted from Figs. S3(a-d) at the wave vectors $(1,0)$ and $(0,1)$, respectively. 
Therefore, the applied uniaxial pressure clearly has an impact on magnetic excitations at $(1,0)$ but
has no observable effect at $(0,1)$, which has weak and featureless energy dependence of isotropic 
$M_a$, $M_b$ and $M_c$ [Fig. S3(h)].

Consistent with the weak scattering at $(0, 1, L)$ observed at 145K, temperature dependence of $M_a$, $M_b$ and $M_c$ at ${\bf Q}=(0, 1)$ is much weaker than that at $(1, 0, L)$ and decreases in intensity at $T_N$ (Fig. S4), 
consistent with the temperature dependence of (0, 1, 1) in detwinned {\BFA} measured 
with unpolarized neutron scattering \cite{xylu18}.

{\bf Uniaxial pressure dependence of the magnetic order and correlations.}
Fig. S5 summarizes the elastic $\theta-2\theta$ scans of $\sigma^{SF}_x$ across ${\bf Q}=(1, 0, L) ~(L=1, 2, 3)$. Similar to the $\theta-2\theta$ scans of $\sigma^{SF}_z$ as described in Fig. 4 of the main text, the scans for $\sigma^{SF}_x$ [Fig. S5(a) and S5(b)] exhibits temperature-independent full-width-at-half-maximum (FWHM) from 40 to 143K [Fig. S5(c)], indicating that the spin-spin correlation length are resolution limited even in the temperature range above $T_N\sim136$ K of unstrained sample. Fig. S5(d) plots the ratio between the scattering intensity at $(1, 0, 1)$ and $(1, 0, 3)$ ($I_{101}/I_{103}$), which is greatly enhanced close to $T_N$. Since $\sigma^{SF}_x=0.16M_a+0.84M_c$ at ${\bf Q} = (1, 0, 1)$ and $0.37M_a+0.37M_c$ at $(1, 0, 3)$, the enhancement of $I_{101}/I_{103}$ is consistent with the emergence of a $c$-axis magnetic moment induced by uniaxial strain. At temperature where $M_c$ is not induced, the ratio $I_{101}/I_{103}=0.16M_a/0.63M_a\times \frac{\sin^22\theta_2}{\sin^22\theta_1}\approx0.5$ (black dashed line in Fig. S5(d), where $\frac{\sin^22\theta_2}{\sin^22\theta_1}$ accounts for the Lorentz factor. The data points of $I_{101}/I_{103}$ in Fig. S5(d) show that $M_c$ is absent at 149K and below 135K but reaches a maximum at 143K close to $T_N$.  The unpolarized data in Fig. 4(f) shows similar behavior. 

In addition to the emergence of $M_c$, it is also important to understand whether $M_c$ forms a new periodicity along $c$-axis. The magnetic structure factor of the three-dimensional antiferromagnetic order of {\BFA} (${\bf k}=(1, 0, 1)$) results in magnetic peaks at $(1,0,L)$ with $L=1, 3, 5...$ and the absence of magnetic scattering at $(1,0,L)$ with $L=0, 2, 4...$. If the induced $M_c$ forms a larger magnetic unit cell along $c$ axis that ensures the presence of (1,0,1) and (1,0,3), one can expect detectable magnetic scattering at L=0, 2. However, the three-point $\theta-2\theta$ across (1, 0, 2) in Fig. S5 shows that the intensity for (1, 0, 2) is smaller than 1/3000 of (1, 0, 3), which rules out this possibility and further confirm our conclusion about the canting-moment picture as shown in Fig. 1 of the main text. 

{\bf Data availability.} The data that support the findings of this study are available
from the corresponding authors on request, and will be available at \cite{ILL}.

\section{Acknowledgements}
The work at BNU is supported by the NSFC under Grant No. 11734002 and 11922402. 
The work at Rice University is supported by the U.S. NSF DMR-1700081 and the Robert A. Welch Foundation Grant No. C-1839 (P.D.). The work at UCB was supported
by the U.S. DOE BES under Contract
No. DE-AC02-05-CH11231 within the Quantum Materials
Program (KC2202).

\section{Author contributions}
P.D. and X.L. conceived the project. P.L., M.L.K., L.T., G.T. and X.L. prepared the
samples. Polarized inelastic neutron scattering experiments at IN22 were carried out by
P.L., L.T., X.L., K.S., J.T.P., Y.L., Y.S., and F. B.  Neutron diffraction measurements
at BT-7 were carried out by M.L.K, Y.S., D.W.T., J.W.L, Y. Z., and R.J.B. 
The entire project was supervised
by P.D. The manuscript was written by P.D., P.L., X.L. and Y.S. All authors
made comments.


\begin{thebibliography}{}

\bibitem{hosono} H. Hosono and K. Kuroki, Physica C, {\bf 514}, 399 (2015).

\bibitem{Johnston} D. C. Johnston, Adv. Phys. {\bf 59}, 803 (2010).

\bibitem{stewart} G. R. Stewart, Rev. Mod. Phys. {\bf 83}, 1589 (2011).

\bibitem{scalapinormp} D. J. Scalapino, Rev. Mod. Phys. {\bf 84}, 1383 (2012).

\bibitem{dai} P. C. Dai, Rev. Mod. Phys. {\bf 87}, 855 (2015).

\bibitem{qhuang} Q. Huang, Y. Qiu, Wei Bao, M. A. Green, J. W. Lynn, Y. C. Gasparovic, T. Wu, G. Wu, X. H. Chen,  Phys. Rev. Lett. {\bf 101}, 257003 (2008).

\bibitem{kim2011} M. G. Kim, R. M. Fernandes, A. Kreyssig, J. W. Kim, A. Thaler, S. L. Bud'ko, P. C. Canfield, R. J. McQueeney, J. Schmalian, A. I. Goldman,  Phys. Rev. B {\bf 83}, 134522 (2011).

\bibitem{RMFernandes2014} R. M. Fernandes, A. V. Chubukov, J. Schmalian, Nat. Phys. {\bf 10}, 97 (2014).

\bibitem{AEBohmerCRP} A. E. B\"{o}hmer, C. Meingast, C. R. Physique {\bf 17} 90 (2016).

\bibitem{fisher} I. R. Fisher, L. Degiorgi, and Z. X. Shen, Rep. Prog. Phys. {\bf 74}, 124506 (2011).

\bibitem{JHChu2010} J.-H. Chu, J. G. Analytis, K. De Greve, P. L. McMahon, Z. Islam, Y. Yamamoto, and I. R. Fisher, Science {\bf 329,} 824 (2010). 

\bibitem{JHChu2012} J.-H. Chu, H.-H. Kuo, J. G. Analytis, and I. R. Fisher, Science {\bf 337,} 710 (2012).

\bibitem{Tanatar10} M. A. Tanatar, E. C. Blomberg, A. Kreyssig, M. G. Kim,
N. Ni, A. Thaler, S. L. Bud'ko, P. C. Canfield, A. I. Goldman, I. I. Mazin, and R. Prozorov, 
Phys. Rev. B {\bf 81}, 184508 (2010). 

\bibitem{Man15} Haoran Man, Xingye Lu, Justin S. Chen, Rui Zhang, Wenliang Zhang, Huiqian Luo, J. Kulda, A. Ivanov, T. Keller, Emilia Morosan, Qimiao Si, and Pengcheng Dai, Phys. Rev. B {\bf 92}, 134521 (2015).

\bibitem{Tam19} David W. Tam, Weiyi Wang, Li Zhang, Yu Song, Rui Zhang, Scott V. Carr, H. C. Walker, Toby G. Perring, D. T. Adroja, and Pengcheng Dai, Phys. Rev. B {\bf 99}, 134519 (2019).

\bibitem{MYi2017} M. Yi, Y. Zhang, Z.-X. Shen, and D. H. Lu, npj Quantum Materials, {\bf 2}, 57 (2017).

\bibitem{Pfau2019} H. Pfau, C. R. Rotundu, J. C. Palmstrom, S. D. Chen, M. Hashimoto, D. Lu, A. F. Kemper, I. R. Fisher, and Z.-X. Shen, Phys. Rev. B {\bf 99}, 035118 (2019).

\bibitem{Watson2019} Matthew D. Watson, Pavel Dudin, Luke C. Rhodes, Daniil V. Evtushinsky, Hideaki Iwasawa, Saicharan Aswartham, Sabine Wurmehl, Bernd B$\rm \ddot{u}$chner, Moritz Hoesch, Timur K. Kim, npj Quantum Materials {\bf 4}, 36 (2019).

\bibitem{Lu14} Xingye Lu, J. T. Park, R. Zhang, H. Luo, A. H. Nevidomskyy, Q. Si, and Pengcheng Dai, Science {\bf 345}, 657 (2014).

\bibitem{MQHe} Mingquan He, Liran Wang, Felix Ahn, Fr$\rm \acute{e}$d$\rm \acute{e}$ric Hardy, 
Thomas Wolf, Peter Adelmann, J$\rm \ddot{o}$rg Schmalian, Ilya Eremin, 
and Christoph Meingast, Nat. Comm. {\bf 8}, 504 (2017).

\bibitem{xylu18}  Xingye Lu, D. D. Scherer, D. W. Tam, W. Zhang, R. Zhang, H. Luo, L. W. Harriger, H. C. Walker, D. T. Adroja, B. M. Andersen, and Pengcheng Dai, Phys. Rev. Lett. {\bf 121}, 067002 (2018).

\bibitem{Dhital2012} C. Dhital, Z. Yamani, W. Tian, J. Zeretsky, A. S. Sefat, Z. Wang, R. J. Birgeneau, and S. D. Wilson, Phys. Rev. Lett. {\bf 108}, 087001 (2012).

\bibitem{Dhital2014} C. Dhital, T. Hogan, Z. Yamani, R. J. Birgeneau, W. Tian, M. Matsuda, A. S. Sefat, Z. Wang, and S. D. Wilson, Phys. Rev. B 89, 214404 (2014).

\bibitem{YSong2013} Y. Song, S. V. Carr, X. Y. Lu, C. L. Zhang, Z. C. Sims, N. F. Luttrell, S. X. Chi, Y. Zhao, J. W. Lynn, and Pengcheng Dai, Phys. Rev. B 87, 184511 (2013).

\bibitem{Tam2017} David W. Tam, Yu Song, Haoran Man, Sky C. Cheung, Zhiping Yin, Xingye Lu, Weiyi Wang, Benjamin A. Frandsen, Lian Liu, Zizhou Gong, Takashi U. Ito, Yipeng Cai, Murray N. Wilson, Shengli Guo, Keisuke Koshiishi, Wei Tian, Bassam Hitti, Alexandre Ivanov, Yang Zhao, Jeffrey W. Lynn, Graeme M. Luke, Tom Berlijn, Thomas A. Maier, Yasutomo J. Uemura, and Pengcheng Dai, 
Phys. Rev. B {\bf 95}, 060505(R) (2017). 

\bibitem{Kissikov} T. Kissikov, R. Sarkar, M. Lawson, B. T. Bush, E. I. Timmons, M. A. Tanatar, R. Prozorov, S. L. Bud'ko, P. C. Canfield, R. M. Fernandes, and N. J. Curro, 
Nat. Comm. {\bf 9}, 1058 (2018).

\bibitem{Lu2016} Xingye Lu, Kuo-Feng Tseng, T. Keller, Wenliang Zhang, Ding Hu, Yu Song, Haoran Man, J. T. Park, Huiqian Luo, Shiliang Li, Andriy H. Nevidomskyy, and Pengcheng Dai, Phys. Rev. B {\bf 93}, 134519 (2016).  


\bibitem{collins} M. F. Collins, {\it Magnetic Critical Scattering} (Oxford University Press, New York, 1989).

\bibitem{birgeneau70} R. J. Birgeneau, H. J. Guggenheim, and G. Shirane, Phys. Rev. B {\bf 1}, 2211 (1970). 

\bibitem{birgeneau75} J. Als-Nielsen, R. J. Birgeneau, H. J. Guggenheim, and G. Shirane, Phys. Rev. B {\bf 12}, 4963 (1975).

\bibitem{MnF2} M. P. Schulhof, R. Nathans, P. Heller, and A. Linz, Phys. Rev. B {\bf 4}, 2254 (1971).

\bibitem{tseng16} K. F. Tseng,  T. Keller,  A. C. Walters,  R. J. Birgeneau, and B. Keimer, Phys. Rev. B {\bf 94}, 014424 (2016).

\bibitem{Wilson10} S. D. Wilson, Z. Yamani, C. R. Rotundu, B. Freelon, P. N. Valdivia, E. Bourret-Courchesne, J. W. Lynn, Songxue Chi, Tao Hong, and R. J. Birgeneau, Phys. Rev. B {\bf 82}, 144502 (2010).


\bibitem{yuli} Yu Li, Weiyi Wang, Yu Song, Haoran Man, Xingye Lu, Fr$\rm \acute{e}$d$\rm \acute{e}$ric Bourdarot, and Pengcheng Dai, Phys. Rev. B {\bf 96}, 020404(R) (2017).

\bibitem{Lipscombe} O. J. Lipscombe, Leland W. Harriger, P. G. Freeman,
M. Enderle, Chenglin Zhang, Miaoying Wang, Takeshi Egami, Jiangping
Hu, Tao Xiang, M. R. Norman, and Pengcheng Dai, Phys. Rev. B \textbf{82},
064515 (2010).

\bibitem{NQureshi2014} N. Qureshi, C. H. Lee, K. Kihou, K. Schmalzl,
P. Steffens, and M. Braden, Phys. Rev. B \textbf{90}, 100502 (2014).


\bibitem{CWangPRX} Chong Wang, Rui Zhang, Fa Wang, Huiqian Luo, L.
P. Regnault, Pengcheng Dai, and Yuan Li, Phys. Rev. X \textbf{3},
041036 (2013).

\bibitem{YS1} Yu Song, Louis-Pierre Regnault, Chenglin Zhang, Guotai Tan, Scott V. Carr, Songxue Chi, A.D. Christianson, Tao Xiang, and Pengcheng Dai, Phys. Rev. B \textbf{88}, 134512 (2013).

\bibitem{CL1} Chenglin Zhang, Yu Song, L.-P. Regnault, Yixi Su, M. Enderle, J. Kulda, Guotai Tan, Zachary C. Sims, Takeshi Egami, Qimiao Si, and Pengcheng Dai, Phys. Rev. B \textbf{90}, 140502 (2014).

\bibitem{lynn1} J. W. Lynn, Y. Chen, S. Chang, Y. Zhao, S. Chi, W. Ratcliff, II, B. G. Ueland, and R. W. Erwin, Journal of Research of NIST {\bf 117}, 61 (2012).

\bibitem{lynn2}  W. C. Chen, G. Armstrong, Y. Chen, B. Collett, R. Erwin, T. R. Gentile, G. L. Jones, J. W. Lynn, S. McKenney, and J. E. Steinberg, Physica B {\bf 397}, 168 (2007).

\bibitem{SI} See supplementary information for additional data and analysis. 


\bibitem{Avci2014} S. Avci, O. Chmaissem, J.M. Allred, S. Rosenkranz, I. Eremin, A.V. Chubukov, D.E. Bugaris, D.Y. Chung, M.G. Kanatzidis, J.-P Castellan, J.A. Schlueter, H. Claus, D.D. Khalyavin, P. Manuel, A. Daoud-Aladine, and R. Osborn, Nat. Comm. {\bf 5}, 3845 (2014).

\bibitem{Allred2016}  J. M. Allred, K. M. Taddei, D. E. Bugaris, M. J. Krogstad,
S. H. Lapidus, D. Y. Chung, H. Claus, M. G. Kanatzidis, D. E.
Brown, J. Kang, R. M. Fernandes, I. Eremin, S. Rosenkranz,
O. Chmaissem, and R. Osborn, Nat. Phys. {\bf 12}, 493 (2016).

\bibitem{Scherer} Daniel D. Scherer and Brian M. Andersen, Phys. Rev. Lett. {\bf 121}, 037205 (2018).

\bibitem{Nakajima} M. Nakajima, T. Liang, S. Ishida, Y. Tomioka, K. Kihou, C. H. Lee, A. Iyo, H. Eisaki, T. Kakeshita, T. Ito, and S. Uchida, PNAS {\bf 108}, 12238 (2011).  

\bibitem{Ren2015} Xiao Ren, Lian Duan, Yuwen Hu, Jiarui Li, Rui Zhang, Huiqian Luo, Pengcheng Dai, and Yuan Li, Phys. Rev. Lett. {\bf 115}, 197002 (2015).

\bibitem{Baum} A. Baum, Ying Li, M. Tomi$\rm \acute{c}$, N. Lazarevi$\rm \acute{c}$, D. Jost, F. L$\rm \ddot{o}$ffler, B. Muschler, T. B$\rm \ddot{o}$hm, J.-H. Chu, I. R. Fisher, R. Valenti, I. I. Mazin, and R. Hackl, Phys. Rev. B {\bf 98}, 075113 (2018).

\bibitem{ILL} All raw data from ILL will be published at DOI:10.5291/ILL-DATA.4-02-531, and from NCNR will be provided upon request. 

\bibitem{tas} Gen Shirane, Stephen M. Shapiro, and John M. Tranquada, Neutron Scattering with a Triple-Axis Spectrometer, Cambridge University Press 2004, p.170.

\end{thebibliography}
\end{document}